\documentclass[10pt,twocolumn]{article}
\usepackage[margin=0.55in,columnsep=0.22in]{geometry}
\usepackage[T1]{fontenc}
\usepackage{lmodern}
\usepackage{microtype}
\usepackage{booktabs}
\usepackage{array}
\usepackage{tabularx}
\usepackage{enumitem}
\usepackage{titlesec}
\usepackage{balance}
\usepackage{xcolor}
\usepackage{amsmath}
\usepackage{amssymb}
\usepackage{tikz}
\usetikzlibrary{arrows.meta,positioning,fit,calc}
\usepackage[numbers,sort&compress]{natbib}
\usepackage{xurl}
\usepackage{hyperref}
\hypersetup{colorlinks=true,linkcolor=blue!55!black,citecolor=blue!55!black,urlcolor=blue!55!black}
\setlist[itemize,enumerate]{nosep,leftmargin=*}
\setlist[description]{style=nextline,font=\normalfont\bfseries,leftmargin=0pt,labelindent=0pt}
\titlespacing*{\section}{0pt}{1.4ex plus .4ex minus .2ex}{.7ex}
\titlespacing*{\subsection}{0pt}{1.0ex plus .3ex minus .2ex}{.45ex}
\newcolumntype{Y}{>{\raggedright\arraybackslash}X}
\newcolumntype{P}[1]{>{\raggedright\arraybackslash}p{#1}}

\newcommand{\AnalysisSourceShaShort}{12792327f33f\ldots 0a4fc382}
\newcommand{\SelectedTasks}{20}
\newcommand{\SelectedCells}{60}
\newcommand{\TerminalCells}{60}
\newcommand{\ScoredCells}{58}
\newcommand{\NullCells}{2}
\newcommand{\BZeroPlanned}{20}
\newcommand{\BZeroScored}{20}
\newcommand{\BZeroNull}{0}
\newcommand{\BZeroSuccesses}{14}
\newcommand{\BZeroRate}{70.0}
\newcommand{\BTwoPlanned}{20}
\newcommand{\BTwoScored}{19}
\newcommand{\BTwoNull}{1}
\newcommand{\BTwoSuccesses}{11}
\newcommand{\BTwoRate}{57.9}
\newcommand{\BThreePlanned}{20}
\newcommand{\BThreeScored}{19}
\newcommand{\BThreeNull}{1}
\newcommand{\BThreeSuccesses}{7}
\newcommand{\BThreeRate}{36.8}
\newcommand{\BThreeBTwoPairs}{18}
\newcommand{\BThreeBTwoDifference}{-0.222}
\newcommand{\BThreeBTwoCILow}{-0.444}
\newcommand{\BThreeBTwoCIHigh}{-0.056}
\newcommand{\BThreeBTwoP}{0.125}
\newcommand{\BThreeOnlyBTwo}{0}
\newcommand{\BTwoOnlyBThree}{4}
\newcommand{\BThreeBTwoTies}{14}
\newcommand{\BThreeBTwoNullLow}{-0.25}
\newcommand{\BThreeBTwoNullHigh}{-0.15}
\newcommand{\BThreeAfterBTwo}{-0.333}
\newcommand{\BThreeBeforeBTwo}{-0.111}
\newcommand{\BThreeBTwoEarly}{-0.222}
\newcommand{\BThreeBTwoLate}{-0.222}
\newcommand{\BThreeBTwoLotoLow}{-0.235}
\newcommand{\BThreeBTwoLotoHigh}{-0.176}
\newcommand{\BThreeBZeroPairs}{19}
\newcommand{\BThreeBZeroDifference}{-0.316}
\newcommand{\BThreeBZeroCILow}{-0.526}
\newcommand{\BThreeBZeroCIHigh}{-0.105}
\newcommand{\BThreeBZeroP}{0.03125}
\newcommand{\BThreeOnlyBZero}{0}
\newcommand{\BZeroOnlyBThree}{6}
\newcommand{\BThreeBZeroTies}{13}
\newcommand{\BThreeBZeroNullLow}{-0.35}
\newcommand{\BThreeBZeroNullHigh}{-0.30}
\newcommand{\BThreeAfterBZero}{-0.333}
\newcommand{\BThreeBeforeBZero}{-0.300}
\newcommand{\BThreeBZeroEarly}{-0.444}
\newcommand{\BThreeBZeroLate}{-0.200}
\newcommand{\BThreeBZeroLotoLow}{-0.333}
\newcommand{\BThreeBZeroLotoHigh}{-0.278}
\newcommand{\BThreeFinalized}{0}
\newcommand{\BThreeClosureDenominator}{19}
\newcommand{\FailureInitialization}{6}
\newcommand{\FailureArtifactReport}{8}
\newcommand{\FailureTransport}{3}
\newcommand{\FailureBootstrap}{2}

\newcommand{\PublicationAvailabilityStatement}{The evaluated treatment was frozen locally at Git commit \texttt{54442e22c3ce483823c9\allowbreak 11dfa8d03a52c85922e6} and was subsequently archived without source modification as \href{https://github.com/amanayayatu-tech/loop-skill/releases/tag/paper-treatment-v3.3.12}{\texttt{paper-treatment-v3.3.12}}. The versioned paper, source, aggregate evidence, checksums, and exact treatment snapshot are archived at \href{https://doi.org/10.5281/zenodo.21616787}{doi:10.5281/zenodo.21616787}.}

\title{LoopSkill: Evidence-Bound Lifecycle Control and the\\Safety--Liveness Tradeoff in Coding Agents}
\author{
  Zengjian Liu\\Peachy Lab
  \and
  Tommy He\\Peachy Lab
  \and
  ELE Zhao\\Peachy Lab
  \and
  HongLiang Sun\\Peachy Lab
}
\date{}

\begin{document}
\maketitle

\begin{abstract}
Coding agents can edit a repository, run tests, review their own work, and announce completion inside one conversational process.  These capabilities make useful automation possible, but they also collapse the distinction between doing work and authorizing a durable lifecycle transition.  A plausible claim of ``tested'' or ``reviewed'' is not evidence that the claimed artifact, dispatch, state version, and authority still agree.  We present LoopSkill, an evidence-bound lifecycle control plane that makes a schema-validated State Gateway the sole canonical writer; binds routes, artifacts, reports, reviews, recovery, and finalization to explicit identities; and fails closed when those identities disagree.

We evaluate the locally frozen integration v3.3.12 under a common 1,800-second budget on an outcome-blind selected subset of SWE-bench Lite tasks.  The original 140-cell confirmatory plan was administratively truncated after interim outcomes were visible; the resulting \SelectedTasks-task, \SelectedCells-cell study is therefore exploratory and nonconfirmatory.  All \TerminalCells{} cells were terminal, \ScoredCells{} were scored, and \NullCells{} were apparatus null.  Blind immutable-artifact success was \BThreeSuccesses/\BThreeScored{} for LoopSkill (B3), \BTwoSuccesses/\BTwoScored{} for implementation plus independent review (B2), and \BZeroSuccesses/\BZeroScored{} for an ordinary coding agent (B0).  Across \BThreeBTwoPairs{} complete B3--B2 pairs, the descriptive paired risk difference was \BThreeBTwoDifference{} (paired-bootstrap 95\% interval $[\BThreeBTwoCILow,\BThreeBTwoCIHigh]$; exact McNemar $p=\BThreeBTwoP$).  None of \BThreeClosureDenominator{} scored B3 workflows reached finalization acknowledgement.  Source inspection establishes that the implementation contains identity-bound, fail-closed checks, while aggregate terminal labels show where observed executions were rejected; the available aggregate evidence does not estimate a false-accept rate or prove no-advance semantics for every rejection.  The combination exposes a severe safety--liveness and fixed-budget cost tradeoff, but does not establish long-horizon efficacy, isolated-agent performance, or comparative superiority.
\end{abstract}

\section{Introduction}

Consider a coding task that spans a Controller, an implementation Worker, a Reviewer, a local test process, and an automation heartbeat.  The Worker returns a patch and says that tests passed.  The Reviewer comments on a patch with the same filename.  Meanwhile, a retry has created a second task, the source tree has changed, or the report references a digest from a prior snapshot.  A conversational trace can make this sequence look coherent even though no single durable record proves that the reviewed bytes are the dispatched bytes, that the report belongs to the active route, or that the terminal authority was still valid when completion was committed.

This problem is not solved by making the model more fluent.  It is a systems problem at the boundary between probabilistic actors and durable effects.  Tool-using agents such as ReAct and Toolformer interleave reasoning with environment actions~\cite{yao2023react,schick2023toolformer}.  Coding-agent systems provide repository navigation, execution environments, and agent-computer interfaces~\cite{yang2024sweagent,wang2025openhands,zhang2024autocoderover}.  Durable workflow systems preserve progress across process failure~\cite{burckhardt2019rsm,burckhardt2021netherite,temporal2026docs}.  Provenance systems bind artifacts to producing steps~\cite{w3c2013prov,torresarias2019intoto}.  Yet a coding-agent lifecycle needs these ideas to compose: state advancement must be atomic, evidence must be bound to the current route and artifact, lost observations must not authorize duplicate work, and internal completion must remain separate from release or publication authority.

Loop engineering frames autonomous coding work as a bounded loop with a goal, verification step, memory, and stopping rule~\cite{macedo2026loops}.  Evidence-gated lifecycle control goes further by treating an agent output as a claim that must satisfy a current gate~\cite{huang2026prooforstop}.  Commit-time authorization similarly emphasizes that authority observed earlier need not remain valid when a durable effect is committed~\cite{santosgrueiro2026authority}.  LoopSkill operationalizes these concerns as a transactional control plane for multi-role coding work.  Its State Gateway is the only component allowed to advance canonical state.  A route is a durable outbox record, not merely a message.  A formal report is admitted only when state, Goal, dispatch, task, artifact, and report identities agree.  A lost response is reconciled against the original outbox.  Finalization requires a separately observed pause of the registered heartbeat.

This design intentionally prefers rejection under ambiguity.  That preference creates a second systems question: can the protocol remain live?  A control plane that rejects every request would prevent invalid advancement but would be useless.  Conversely, a highly permissive agent loop may finish often while silently accepting stale or foreign evidence.  LoopSkill therefore has to be evaluated along four distinct axes:

\begin{description}
  \item[Safety] Does the control plane prevent unsupported or identity-inconsistent promotion?
  \item[Liveness] Can a legitimate execution traverse routing, reporting, review, and finalization within its operating budget?
  \item[Efficacy] Does the frozen artifact solve the coding task under a treatment-external blind Oracle?
  \item[Cost] What time and completion opportunities are consumed by the additional control protocol?
\end{description}

The separation is essential.  A correct patch with a blocked formal report is an efficacy success and a liveness failure.  A clean finalization receipt does not prove semantic correctness.  A schema rejection can demonstrate one boundary behavior without establishing a population-level safety rate.  This paper reports each outcome at the level the evidence permits.

The evaluation uses a shared 30-minute budget because the experiment compares complete end-to-end systems rather than granting B3 free control-plane time.  This makes orchestration overhead part of the treatment.  It also sharply limits external validity: the study evaluates fixed-budget repository repair, not hours-long, interruption-rich work.  The original experiment targeted 140 cells.  After interim outcomes were visible, execution was administratively truncated to a previously selected 20-task subset.  The locked analysis is consequently exploratory and must not be read as a confirmatory superiority or inferiority test.

This paper makes four contributions:

\begin{enumerate}
  \item It specifies an authority-separated lifecycle architecture in which a schema-validated Gateway is the sole canonical writer and all advancement evidence is bound to current state, route, artifact, and role identities.
  \item It describes the implemented v3.3.12 protocol for durable outbox routing, runtime-mediated artifact/report staging, original-outbox recovery, independent review, heartbeat-gated finalization, and fail-closed rejection.
  \item It reports an exploratory three-condition evaluation that measures blind artifact efficacy independently from workflow closure and exposes the fixed-budget safety--liveness tradeoff rather than hiding it behind a single success metric.
  \item It provides a post-hoc descriptive failure taxonomy and a machine-derived evidence interface that keeps every reported denominator, pair count, interval, null bound, and closure count synchronized with the frozen public analysis.
\end{enumerate}

The bounded conclusion is not that LoopSkill is superior, nor that lifecycle control is unnecessary.  The evaluated integration enforced meaningful evidence boundaries but did not remain live enough to achieve competitive observed efficacy in this setting.  That negative tension is the central empirical result and a concrete design agenda: deterministic protocol adaptation should move from language-model reasoning into the runtime while semantic implementation remains with the model.

\section{Background and Problem Statement}

\subsection{From tool use to durable lifecycle authority}

Agent benchmarks increasingly evaluate multi-step interaction rather than isolated generation.  AgentBench, GAIA, OSWorld, AgentBoard, and MLAgentBench cover tool use, planning, computer interaction, and long-running experimentation~\cite{liu2024agentbench,mialon2024gaia,xie2024osworld,ma2024agentboard,huang2024mlagentbench}.  SWE-bench focuses evaluation on real repository issues and executable tests~\cite{jimenez2024swebench}.  These benchmarks answer whether an agent reaches a task outcome.  They generally do not require every intermediate lifecycle claim to be admitted by an identity-bound state transition.

Coding-agent loops introduce at least three kinds of state.  First, \emph{workspace state} consists of source files, tests, and generated artifacts.  Second, \emph{conversational state} consists of messages and model context.  Third, \emph{canonical lifecycle state} says which Goal is active, which route was sent, which result was acknowledged, and whether review and finalization are complete.  Treating any one of these as a substitute for the others creates ambiguity.  A source diff can exist without a valid route acknowledgement; a task can say ``complete'' without an immutable diff; a canonical route can remain pending after the Worker actually executed.

LoopSkill's problem statement is therefore:

\begin{quote}
Given fallible model roles, fallible transport, and durable workspace effects, accept a lifecycle transition only when the exact current identities and required evidence agree, while recovering prior side effects without duplicating product work and preserving enough liveness to complete useful tasks.
\end{quote}

The problem resembles state-machine replication in its insistence on ordered transitions~\cite{schneider1990statemachine,lamport1998parliament}, sagas in its decomposition of a long transaction~\cite{garcia1987sagas}, and workflow patterns in its explicit control flow~\cite{vanderaalst2003workflow}.  It differs in two respects.  The transition proposer is a language model that may produce syntactically plausible but schema-invalid adaptations, and semantic correctness is evaluated later by an external Oracle rather than proved by the control plane.

\subsection{Four properties, not one score}

Let $S_v$ be canonical state at version $v$, $r$ a route/outbox, $d$ its dispatch, $a$ an immutable artifact identity, $q$ a staged report, and $h$ an external heartbeat observation.  Let $\mathsf{Accept}(m,S_v)$ denote that mutation $m$ advances canonical state.

\paragraph{Safety.}
For each accepted mutation, the required identities must be current and mutually bound:
\[
\begin{aligned}
\mathsf{Accept}(m,S_v) \Rightarrow{}& \mathsf{Current}(v)\\
  &{}\land \mathsf{Authorized}(m) \land \mathsf{Bound}(r,d,a,q,h).
\end{aligned}
\]
A rejected mutation must leave canonical semantics unchanged.  This is a transition-safety property, not a proof that artifact $a$ is correct.

\paragraph{Liveness.}
Under valid inputs and available external services, a legitimate route should eventually reach an acknowledged result and, after required review, finalization.  The implementation is fail-closed and budgeted, so this is an empirical operational objective rather than a proved guarantee.  We observe it through the proportion of eligible B3 executions that reach \texttt{FINALIZATION\_ACKED}.

\paragraph{Efficacy.}
Efficacy is $Y(a)\in\{0,1\}$ from a treatment-external blind Oracle applied once to the frozen artifact.  The Oracle does not inspect lifecycle rhetoric.  Apparatus failures that prevent a valid classification remain null rather than becoming product failures or successes.

\paragraph{Cost.}
Cost includes the common wall-clock budget consumed by protocol actions and the opportunity cost of failing to produce a scorable or finalized outcome.  The locked public analysis does not provide a complete provider-token or monetary ledger, so we make no quantitative token-cost claim.  Within the experiment, however, B3's routing, report, review, recovery, and finalization time is part of B3, not removable overhead.

\subsection{Why fail-closed behavior is insufficient}

A failure detector can appear strong by refusing progress.  This is the classic safety--liveness tension: preventing a bad state says nothing about reaching a good terminal state.  Software supply-chain systems make provenance and authorized roles explicit~\cite{torresarias2019intoto,slsa2026spec,cappos2010tuf}; durable execution systems recover work after interruption~\cite{temporal2026docs}; a coding-agent control plane needs both.  LoopSkill's experimental question is not merely whether invalid evidence is rejected, but whether the complete integration can preserve artifact efficacy and terminal progress under an equal budget.

\section{System and Threat Model}

\subsection{Actors and authority}

Figure~\ref{fig:architecture} summarizes the architecture.  Authority is separated by action rather than by a role's self-description.  Table~\ref{tab:authority} states the critical boundaries.

\begin{figure*}[t]
  \centering
  \resizebox{\linewidth}{!}{\begin{tikzpicture}[
  font=\scriptsize,
  node distance=5mm and 7mm,
  role/.style={draw, rounded corners, align=center, minimum height=7mm, fill=blue!5},
  gate/.style={draw, very thick, rounded corners, align=center, minimum height=9mm, fill=orange!12},
  runtime/.style={draw, dashed, rounded corners, align=center, minimum height=7mm, fill=gray!8},
  authority/.style={draw, rounded corners, align=center, minimum height=7mm, fill=green!8},
  flow/.style={-Latex, thick},
  evidence/.style={-Latex, thick, dashed},
]
  \node[role] (controller) {Controller\\route and reconcile};
  \node[gate, right=of controller, minimum width=28mm] (gateway) {State Gateway\\sole canonical writer};
  \node[role, right=of gateway] (worker) {Worker\\bounded edits};
  \node[role, below=of worker] (reviewer) {Reviewer\\read-only decision};
  \node[runtime, below=of gateway] (codec) {Runtime codec\\materialize, capture, stage};
  \node[authority, below=of controller] (heartbeat) {External heartbeat\\pause and readback};
  \node[authority, above=of gateway] (neutral) {Neutral experiment authority\\schedule, capture, Oracle, analysis};

  \draw[flow] (controller) -- node[above]{CAS request} (gateway);
  \draw[flow] (gateway) -- node[above]{bound dispatch} (worker);
  \draw[flow] (worker) -- (codec);
  \draw[flow] (codec) -- node[left]{stage} (gateway);
  \draw[flow] (gateway) -- (reviewer);
  \draw[flow] (reviewer) -- (codec);
  \draw[evidence] (heartbeat) -- node[left]{terminal witness} (gateway);
  \draw[evidence] (neutral) -- node[right]{outside treatment} (gateway);

  \node[below=5mm of codec, align=center, text width=.92\linewidth] (boundary) {Every accepted transition binds current state version, Goal, lease, route/outbox, dispatch, task, artifact, report, and role.\\The neutral authority never enters B3's canonical state and B3 never controls scheduling or scoring.};
\end{tikzpicture}}
  \caption{LoopSkill authority architecture.  Solid arrows are lifecycle operations; dashed arrows are evidence or external observations.  The experiment authority remains outside the B3 treatment.}
  \label{fig:architecture}
\end{figure*}

\begin{table*}[t]
\centering
\caption{Actors, powers, and explicit non-authorities.}
\label{tab:authority}
\small
\begin{tabularx}{\textwidth}{@{}P{25mm}YY@{}}
\toprule
Actor & Authorized action & Insufficient or forbidden action \\
\midrule
Controller & Select an eligible Goal, acquire one lease, prepare one route, reconcile its original outbox & Edit product files, fabricate a Worker result, or write canonical state directly \\
Worker & Edit the dispatched paths; capture and stage its exact result through the runtime & Create routes, broaden scope, self-review, or hand-write canonical state \\
Reviewer & Inspect the exact Worker artifact and stage a role-specific read-only decision & Repair the artifact it judges or reuse the Worker dispatch \\
Runtime codec & Materialize dispatches, capture complete diffs, and stage immutable reports/evidence & Decide lifecycle promotion or become a second state writer \\
State Gateway & Validate and atomically apply authorized state transitions & Judge semantic patch correctness or perform App-side sends \\
Heartbeat authority & Provide independently observed pause/readback evidence & Declare artifact correctness, release, or publication \\
Neutral experiment authority & Select, schedule, freeze artifacts, run the Oracle, classify, and analyze & Enter the B3 treatment or change its protocol for a result \\
Human author & Approve authorship, public evidence, release, and submission & None of these permissions is inferred from internal finalization \\
\bottomrule
\end{tabularx}
\end{table*}

The Controller is a coordinator, not a privileged writer.  The Worker has repository edit authority only within the immutable Goal.  The Reviewer is independent in the procedural sense that it receives a distinct dispatch and cannot edit the Worker artifact while judging it.  The runtime codec is trusted for bounded mechanical transformations but not for lifecycle policy.  Only the Gateway mutates canonical state.  The external heartbeat authority reports an operational fact.  Public release, submission, and authorship remain outside the protocol.

\subsection{Protected identities}

Each accepted result is tied to a chain of identifiers rather than a filename or prose claim.  The chain includes the state version, Goal and milestone, lease, route/outbox, payload digest, target task, model configuration, workspace root, base revision, artifact snapshot, complete-diff reference, validation evidence, report bytes, review dispatch, runtime, and terminal predecessor where applicable.  Identity equality is structural: a semantically similar report with the wrong digest is foreign evidence.

Three invariants follow.

\begin{description}
  \item[I1: sole canonical writer.] Only a valid Gateway mutation may change canonical lifecycle state.
  \item[I2: identity-bound promotion.] A report or review can advance only its current route and exact artifact; stale, replayed, or foreign evidence has zero advancement effect.
  \item[I3: non-escalating terminality.] \texttt{FINALIZATION\_ACKED} closes an internal lifecycle only.  It never implies release, merge, publication, peer review, or scientific validity.
\end{description}

\subsection{Fault and adversary model}

We consider stale state versions, wrong Goal or task identities, wrong artifacts, replay and duplicate delivery, malformed requests, unauthorized scope expansion, lost stdout, delayed App acknowledgements, process interruption, network degradation, model/configuration drift, successor confusion, and premature finalization.  We also consider authority collapse: a Worker may attempt to self-approve, a Reviewer may edit, or a Controller may treat observation as canonical state.

The required response to an unproven transition is rejection with unchanged canonical semantics and retained negative evidence.  A transport error is not permission to send twice.  A missing report response is not permission to rerun implementation.  Recovery must refer to the original outbox and staged bytes.  The model does not defend a fully compromised host, provider, or kernel; it does not offer cryptographic provider identity; and it cannot derive semantic correctness from a receipt.

\section{LoopSkill Design}

\subsection{Schema-validated canonical state}

LoopSkill stores an explicit lifecycle state whose schema fixes required fields, permitted enumerations, and cross-object identities.  A Gateway request carries an operation, caller role, expected version, idempotency identity, and the operation-specific evidence.  Validation proceeds before commit.  If the request is stale, malformed, unauthorized, or inconsistent with the current predecessor, the Gateway rejects it without advancing state.  If it is valid, the state update and event are committed as one canonical transition.

This is a compare-and-set discipline applied to agent work.  It eliminates several ambiguous shortcuts: editing a Markdown status file is not a state transition; a Worker message is not an acknowledged result; a Controller's observation of a task is not proof that the registered route reached it.  The Gateway is deliberately not an execution engine.  It checks admissibility and persists state, while App interactions and repository edits remain external side effects whose evidence is brought back to the gate.

\subsection{Atomic route preparation and durable outbox}

A route begins with \texttt{PREPARE\_ROUTE}.  Under the current lease and state version, the Gateway derives an outbox identity that binds the Goal, role, target, payload digest, and send budget.  The Controller materializes that payload and performs one real App send.  \texttt{RECORD\_ROUTE\_SENT} records the returned target identity and observation against the same outbox.

Separating preparation from send makes ambiguity visible.  If an App call returns a generic error but creates the target asynchronously, the Controller cannot safely assume failure and resend.  It must read back the exact-root task set and resolve whether the prepared route already has one matching target, no target, or multiple targets.  Only the first is a safe adoption.  This is analogous to a transactional outbox pattern~\cite{aws2026outbox}: durable intent precedes the external side effect, and later acknowledgement is tied to that intent.

\subsection{Runtime-mediated artifact and report attestation}

The Worker verifies its dispatch before product work.  On completion, the runtime codec captures the complete authorized diff and constructs a stable artifact identity.  A \texttt{STAGE\_REPORT} request contains the outbox, strict report text, status, artifact digest, and any evidence sources.  Each evidence source has a canonical report path, a source path inside the registered worktree but outside control-plane storage, an exact digest, and a closed media type.  The staged report is immutable.

This design distinguishes three digests that conversational systems often conflate: the digest of an individual evidence file, the digest of the complete artifact snapshot, and the digest of the formal report bytes.  The Gateway admits \texttt{ACK\_ROUTE\_RESULT} only when the route, result, report, and artifact identities form the expected chain.  The control plane proves that the admitted report concerns the dispatched artifact; it still does not prove that the patch fixes the issue.

\subsection{Independent review and maker--checker separation}

After a Worker result is acknowledged, the Gateway prepares a distinct review route.  The Reviewer receives the exact immutable Worker artifact and evidence identities.  It may inspect and decide, but it cannot edit the target artifact.  Its report is staged through the same codec under a Reviewer-specific dispatch.  The Controller then acknowledges that review route with the public \texttt{ACK\_ROUTE\_RESULT} operation; the runtime internally projects the accepted Reviewer report into review state.  There is no public \texttt{RECORD\_REVIEW} Gateway operation.  A review acknowledgement carries zero new product artifacts and references the accepted Worker predecessor.

This separation differs from self-refinement approaches in which one model critiques and rewrites its own output~\cite{shinn2023reflexion,madaan2023selfrefine}.  Both are useful mechanisms, but they answer different questions.  LoopSkill's review is a lifecycle gate and provenance boundary, not a claim of statistically independent human judgment.  Model-based judging can be scalable but has known position, verbosity, and self-enhancement biases~\cite{zheng2023llmjudge}; exact artifact binding prevents identity confusion but does not remove judgment error.

\subsection{Original-outbox recovery}

Lost output is common in long tool runs.  LoopSkill does not infer that an absent response means an absent side effect.  \texttt{REPORT\_RECOVERY} reads persistent state for the original outbox and returns the same staged handle and digests.  Transport recovery similarly inspects the one prepared route, its target identity, and bounded App observations.  It never creates a replacement route merely because stdout was lost.

The safety benefit is duplicate-work prevention.  The liveness cost is that ambiguous external state may leave the route unresolved.  LoopSkill bounds natural observations and repair attempts; when ambiguity remains, the heartbeat is paused and the state retains a blocked outbox.  This is safer than silently adopting the wrong task, but it can exhaust a short execution budget.

\subsection{Heartbeat-gated finalization}

\begin{figure*}[t]
  \centering
  \resizebox{\linewidth}{!}{\begin{tikzpicture}[
  font=\scriptsize,
  lifeline/.style={gray!65, densely dotted},
  message/.style={-Latex, thick},
  reject/.style={-Latex, thick, red!65!black},
]
  \foreach \x/\name in {0/Controller,3/Gateway,6/Worker,9/Reviewer,12/Heartbeat} {
    \node[font=\scriptsize\bfseries] at (\x,0) {\name};
    \draw[lifeline] (\x,-.3) -- (\x,-7.9);
  }
  \draw[message] (0,-.8) -- node[above]{PREPARE\_ROUTE} (3,-.8);
  \draw[message] (3,-1.5) -- node[above]{immutable dispatch} (6,-1.5);
  \draw[message] (6,-2.2) -- node[above]{STAGE\_REPORT} (3,-2.2);
  \draw[message] (0,-2.9) -- node[above]{ACK\_ROUTE\_RESULT} (3,-2.9);
  \draw[message] (3,-3.6) -- node[above]{review dispatch} (9,-3.6);
  \draw[message] (9,-4.3) -- node[above]{STAGE\_REPORT} (3,-4.3);
  \draw[message] (0,-5.0) -- node[above]{ACK\_ROUTE\_RESULT} (3,-5.0);
  \draw[message] (0,-5.7) -- node[above]{PREPARE\_FINALIZATION} (3,-5.7);
  \draw[message] (12,-6.4) -- node[above]{pause/readback witness} (3,-6.4);
  \draw[message] (0,-7.1) -- node[above]{ACK\_FINALIZATION} (3,-7.1);
  \node[draw, rounded corners, fill=green!9, align=center] at (3,-7.85) {FINALIZATION\_ACKED};

  \node[draw, rounded corners, fill=red!6, align=left, text width=43mm] (bad) at (9,-7.85)
    {Fail-closed branch: stale version, foreign identity, malformed report, duplicate delivery, or missing terminal witness $\Rightarrow$ reject with no canonical advancement; retain negative evidence.};
  \draw[reject] (3,-2.2) -- (bad.north west);
\end{tikzpicture}}
  \caption{Simplified transaction sequence.  The real schema carries additional identities and evidence.  Any failed precondition preserves the prior canonical state.}
  \label{fig:protocol}
\end{figure*}
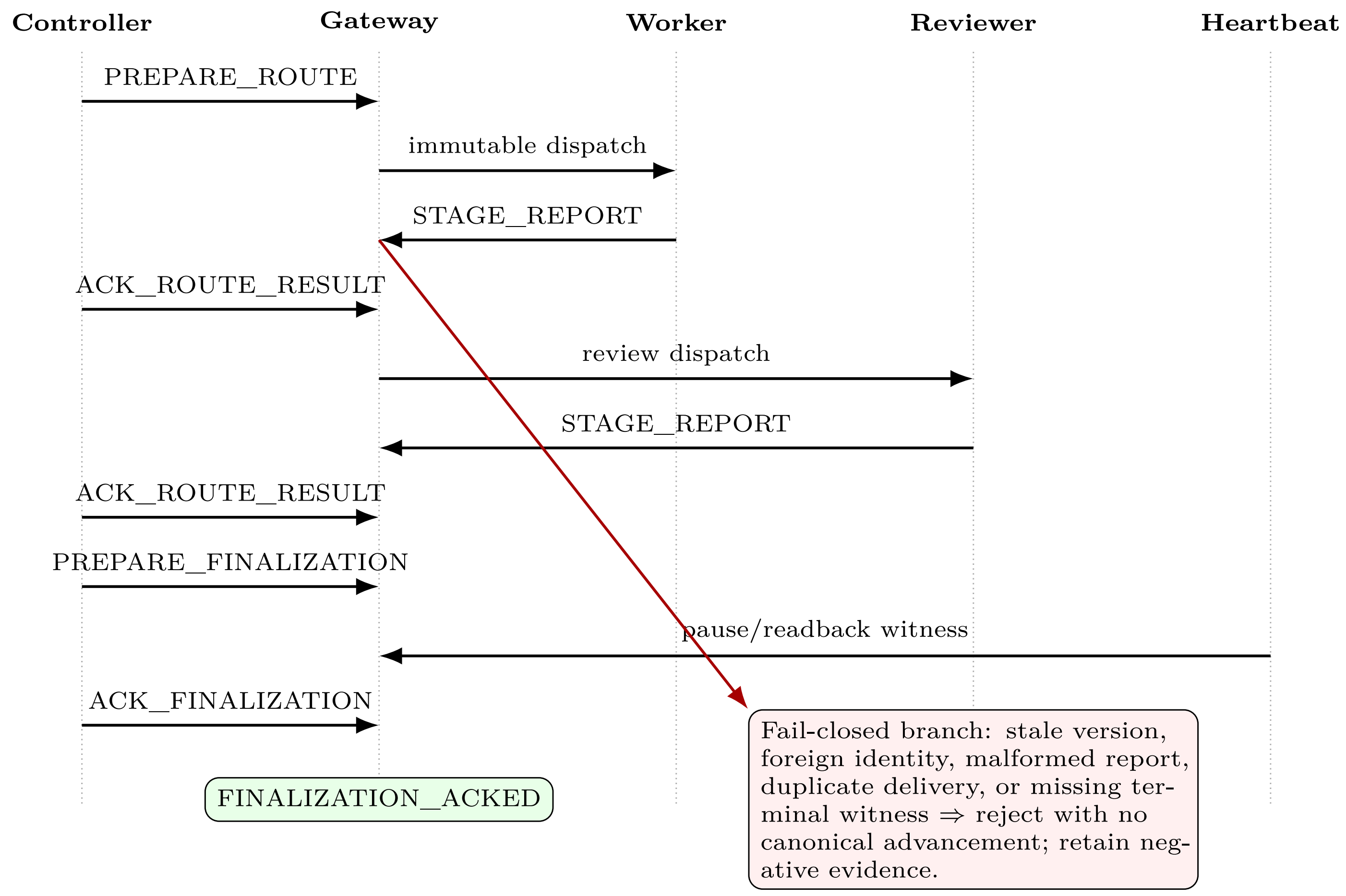

Finalization is a separate transaction rather than a Controller message.  After required results and reviews, \texttt{PREPARE\_FINALIZATION} creates a nonterminal finalization outbox.  The external authority pauses the registered heartbeat and reads back the exact automation identity and status.  Only then may \texttt{ACK\_FINALIZATION} produce \texttt{FINALIZATION\_ACKED}.  Figure~\ref{fig:protocol} shows the happy path and the common fail-closed branch.

The heartbeat serves two purposes.  It prevents a supposedly terminal loop from continuing to route work, and it supplies an external observation that the routing mechanism is quiescent.  It does not upgrade internal completion into publication authority.  A successor begins with a fresh identity and must preserve its predecessor's evidence.

\section{Implementation}

\subsection{Locally frozen identity and code boundary}

The evaluated treatment was the installed local LoopSkill v3.3.12 source at Git commit
\texttt{54442e22c3ce483823c9\allowbreak 11dfa8d03a52c85922e6}.  We use this exact source, schema, and contract as the implementation authority.  The treatment was frozen locally before the experiment; any later public archive is a historical snapshot of those bytes, not a claim that a public release existed during execution.

The implementation has four relevant layers.  First, an Adaptive Pack encodes the Goal graph, roles, budgets, allowed paths, and lifecycle policy.  Second, an MCP service exposes the State Gateway and runtime codec operations to authorized tasks.  Third, the Python runtime implements validation, state transition logic, report/evidence staging, outbox recovery, and heartbeat/finalization checks.  Fourth, a route orchestrator deterministically sequences writer callbacks for \texttt{PREPARE\_ROUTE}, receipt recording through \texttt{RECORD\_ROUTE\_SENT}, and result acknowledgement without itself becoming a state writer.  The caller performs any external App send first and supplies its already obtained receipt; the orchestrator does not execute that external side effect or make it network-atomic.

\subsection{State schema and Gateway dispatcher}

The schema fixes \texttt{state\_gateway\_mode} to \texttt{MCP\_CANONICAL\_WRITER}.  It defines route, role registry, artifact, report, recovery, heartbeat, and finalization records.  The runtime dispatches named mutations including \texttt{PREPARE\_ROUTE}, \texttt{RECORD\_ROUTE\_SENT}, \texttt{ACK\_ROUTE\_RESULT}, \texttt{REPORT\_RECOVERY}, \texttt{PREPARE\_FINALIZATION}, and \texttt{ACK\_FINALIZATION}.  Operation-specific validators run against a copy of the current state; the canonical write occurs only after the mutation completes successfully.

The state version is part of every mutating request.  Idempotent requests may recover their prior result, but a request cannot silently apply to a later state.  Event records preserve both accepted transitions and rejection evidence at the experiment layer.  This design makes state progression auditable without equating a large log with correctness.

\subsection{Runtime codec and immutable staging}

The runtime codec performs deterministic work that should not be delegated to prose interpretation: it validates dispatch transport, verifies registered roots, captures Git or non-Git artifacts, computes digests, validates evidence-source paths and media types, and stages immutable report bytes.  For Git worktrees, complete-diff capture includes the authorized repository delta relative to the frozen base.  Control-plane files are excluded from the task artifact.  Report staging refuses symlinks, paths outside the target worktree, evidence stored under control-plane directories, digest mismatches, and additional schema keys.

These checks are intentionally strict because the artifact identity is later used to connect Worker execution, review, and finalization.  The operational record also shows the downside of allowing a model to assemble a strict request: small naming or digest-selection errors can block a semantically valid patch.  The design lesson is not to weaken the identity check, but to move derivable fields into the runtime so that the model chooses less protocol data.

\subsection{Outbox, heartbeat, and finalization state}

Prepared routes are persisted before App-side delivery.  A sent route retains its outbox until the matching result is acknowledged.  Recovery uses that record to determine whether the correct action is readback, adoption, pause, or refusal.  The role registry prevents an arbitrary task from presenting itself as the intended Worker or Reviewer.

The Pack permits one business heartbeat.  Its deterministic name and prompt identity are registered in canonical state, and later observations must match the same automation.  Transport recovery can require a pause and user-visible notification; finalization requires a matching paused readback.  The terminal receipt binds the finalization outbox and predecessor state version.  No implementation path maps terminality to commit, push, release, or submission.

\subsection{Implementation claims and non-claims}

The code and schema support claims about implemented checks and transition structure.  They do not prove that every possible invalid input is rejected, that the host observation is cryptographically authentic, or that the protocol is live.  The evaluation observes one integration on one host configuration.  The independent public evidence interface therefore distinguishes code-level description, observed boundary behavior, and aggregate empirical outcomes.

\section{Evaluation Method}

\subsection{Research questions}

\begin{description}
  \item[RQ1---principal exploratory efficacy.] In the reduced 20-task population, how does blind task success for LoopSkill B3 differ from implementation plus independent review B2?
  \item[RQ2---secondary exploratory efficacy.] How does B3 differ from the ordinary-agent B0 baseline on the same selected tasks?
  \item[RQ3---workflow liveness.] Separately from artifact correctness, how often does B3 reach \texttt{FINALIZATION\_ACKED}?
  \item[RQ4---operational mechanism.] Which protocol boundary families describe the observed B3 closure failures?
\end{description}

RQ1 and RQ2 are exploratory because of the post-interim administrative truncation.  RQ3 is a direct reliability summary.  RQ4 is a post-hoc descriptive taxonomy, not a preregistered causal analysis.

\subsection{Task population and qualification}

The starting population was SWE-bench Lite, a 300-task subset of SWE-bench's real GitHub issue-resolution benchmark~\cite{jimenez2024swebench}.  Qualification used only public task metadata and apparatus compatibility.  It did not execute hidden tests, read gold patches, or inspect task outcomes.  The procedure produced 61 qualified tasks, 15 permanently ineligible tasks, and zero unresolved tasks.  One qualified task was reserved for apparatus shakedown; the remaining 60 formed the original planned paired population.

Qualification and execution were separate.  A task was not deemed eligible because a treatment succeeded on it.  Image identities, base revisions, public inputs, and zero-hidden-execution receipts were fixed before scoring.  This prevents treatment-specific filtering, although it can still narrow generalizability to tasks compatible with the local frozen apparatus.

\subsection{Original plan and administrative truncation}

The original matrix contained B2 and B3 for all 60 tasks plus B0 for an outcome-blind hash-selected 20-task subset: 140 cells.  During execution, the author shortened the target after interim results were visible, citing time, resource, and construct-alignment concerns.  The reduced population was exactly the pre-existing 20-task B0 subset and all three conditions on those tasks, for \SelectedCells{} selected cells.

The amendment preserved terminal selected cells, launched only missing arms in their original within-task order, and never reran a terminal cell.  Completed cells outside the selected subset were retained as diagnostics but excluded from the reduced estimands.  Reuse of an earlier outcome-blind subset reduces one form of selection freedom, but it does not restore confirmatory validity: the decision to truncate occurred after outcomes were known.  All intervals and $p$-values are descriptive.

\subsection{Conditions}

Table~\ref{tab:conditions} defines the systems compared.  B0 is intentionally simple, reflecting evidence that simpler software-engineering agent pipelines can be competitive and less costly~\cite{xia2025agentless}.  B2 isolates implementation plus just-in-time independent review without LoopSkill's state machine.  B3 evaluates the complete locally frozen integration, not an idealized protocol with failed transitions removed.

\begin{table*}[t]
\centering
\caption{Condition definitions under the common treatment budget.}
\label{tab:conditions}
\small
\begin{tabularx}{\textwidth}{@{}P{14mm}P{42mm}YYY@{}}
\toprule
Arm & Execution & Independent review & LoopSkill control plane & External artifact/Oracle \\
\midrule
B0 & One ordinary coding Worker & No & No & Yes \\
B2 & One coding Worker & One Reviewer; at most one review return & No & Yes \\
B3 & Controller, one Worker, just-in-time Reviewer & Role-bound review route & Gateway, outbox, codec, heartbeat, finalization & Yes \\
\bottomrule
\end{tabularx}
\end{table*}

Each cell received the same frozen public issue statement, a fresh exact-base checkout, and a common 1,800-second treatment budget.  No cell could commit, push, publish, or access hidden tests.  Platform memory was enabled across all executed conditions and held constant.  The estimand is therefore the end-to-end Codex Desktop system under observed memory behavior, not a memory-free isolated model.

\subsection{Neutral authority and experimental boundary}

LoopSkill was the B3 treatment only.  A neutral outer process controlled task selection, condition order, roots, scheduling, project registration, treatment cutoff, artifact capture, Oracle invocation, outcome classification, analysis, and manuscript claims.  B3 could not observe Oracle results or decide which cell ran next.  Figure~\ref{fig:evaluation-boundary} shows the boundary.

\begin{figure*}[t]
  \centering
  \resizebox{\linewidth}{!}{\begin{tikzpicture}[
  font=\scriptsize,
  box/.style={draw, rounded corners, align=center, minimum height=8mm},
  arrow/.style={-Latex, thick},
]
  \node[box, fill=gray!8] (input) {Frozen public task\\exact-base checkout};
  \node[box, right=6mm of input, fill=blue!7] (arms) {Common 1,800-s budget\\B0 ordinary agent\\B2 agent + review\\B3 LoopSkill};
  \node[box, right=6mm of arms, fill=orange!10] (freeze) {Treatment ends\\external immutable\\artifact capture};
  \node[box, right=6mm of freeze, fill=green!8] (oracle) {Blind Oracle once\\network disabled\\binary 0/1 or null};
  \draw[arrow] (input) -- (arms);
  \draw[arrow] (arms) -- (freeze);
  \draw[arrow] (freeze) -- (oracle);

  \node[draw, dashed, rounded corners, fit=(arms), inner sep=4mm, label={[font=\scriptsize]above:treatment boundary}] {};
  \node[below=8mm of freeze, align=center, text width=145mm] (boundarynote)
    {Only B3 uses the Gateway, role graph, heartbeat, report staging, and finalization. Selection, order, roots, artifact capture, private test application, classification, analysis, and paper claims remain treatment-external.};
\end{tikzpicture}}
  \caption{Evaluation boundary.  Artifact correctness is measured after treatment write access ends.  Workflow closure is a separate B3 product-reliability outcome.}
  \label{fig:evaluation-boundary}
\end{figure*}

The original matrix fixed distinct root paths, deterministic launch order, no more than three concurrent lanes, and serial execution within a task.  Project registration occurred before the treatment clock and was machine-bound to root, base commit, clean status, and frozen input.  Missing, duplicate, or mismatched project identities failed closed.

\subsection{Immutable artifact and blind Oracle}

After treatment write access ended, the neutral process captured a binary Git diff relative to the frozen base or an explicit empty artifact.  Control-plane files were excluded.  Capture was required to leave the real index and source status unchanged.  The artifact was then made immutable for scoring.

The common Oracle ran once in the frozen task image with networking disabled.  It restored authoritative test paths, applied private evaluation material, executed prespecified fail-to-pass and pass-to-pass targets, and emitted a binary classification only when identity mapping and target execution were complete.  A valid 0 or 1 was terminal.  An independently established apparatus failure was null.  The design prohibited selective reruns and prohibited sending Oracle outcomes back to treatment tasks.

This artifact gate is why workflow closure cannot substitute for task efficacy.  The Oracle scores bytes, not claims.  Conversely, a passing patch does not retroactively turn an invalid state transition into valid closure.

\subsection{Outcomes and estimands}

The principal descriptive estimand is the mean within-task difference $Y_{B3}-Y_{B2}$ among valid complete pairs in the reduced population.  B3--B0 is secondary and exploratory.  We report paired $2\times2$ counts, paired risk differences, 100,000-replicate paired-bootstrap percentile intervals with locked seeds, exact two-sided McNemar values, worst/best null bounds over all 20 tasks, condition-order strata, launch-order diagnostics, and leave-one-task-out ranges.

B3 workflow liveness is the number of scored B3 cells reaching \texttt{FINALIZATION\_ACKED}.  We do not combine it with Oracle success.  The post-hoc operational taxonomy groups the exact aggregate terminal-reason inventory into four mechanism-oriented families and verifies that every non-null reason maps exactly once.

\subsection{Machine-derived result interface}

To prevent manuscript drift, a paper-local generator reads exactly one frozen public analysis JSON whose SHA-256 is \texttt{\AnalysisSourceShaShort}.  It asserts the status, arm totals, paired counts, intervals, null bounds, closure denominator, and one-to-one failure mapping before emitting LaTeX macros.  The manuscript tables and figures consume those macros.  A check mode recomputes expected bytes without modifying the experiment.  The generator is prohibited from reading raw Oracle material, cell artifacts, or private Codex state.

\section{Results}

\subsection{Cell accounting and arm outcomes}

All \TerminalCells{} selected cells reached a unique terminal state.  \ScoredCells{} produced a valid blind Oracle classification; \NullCells{} remained apparatus null without retry or reclassification.  The B2 null followed prior-solution memory exposure, and the B3 null followed private-material exposure.  These reasons are reported only at the already-approved aggregate level.

\begin{table}[t]
\centering
\caption{Exploratory outcomes in the administratively truncated population.  Success rates use scored-cell denominators.}
\label{tab:arm-results-v2}
\begin{tabular}{lrrrrr}
\toprule
Condition & Planned & Scored & Null & Success & Rate \\
\midrule
B0 & \BZeroPlanned & \BZeroScored & \BZeroNull & \BZeroSuccesses & \BZeroRate\% \\
B2 & \BTwoPlanned & \BTwoScored & \BTwoNull & \BTwoSuccesses & \BTwoRate\% \\
B3 & \BThreePlanned & \BThreeScored & \BThreeNull & \BThreeSuccesses & \BThreeRate\% \\
\bottomrule
\end{tabular}
\end{table}

Observed artifact success was \BZeroSuccesses/\BZeroScored{} for B0, \BTwoSuccesses/\BTwoScored{} for B2, and \BThreeSuccesses/\BThreeScored{} for B3 (Table~\ref{tab:arm-results-v2}).  These marginal rates do not account for pairing or nulls, so the principal result uses complete B3--B2 task pairs.

\subsection{Principal B3--B2 contrast}

There were \BThreeBTwoPairs{} valid pairs.  B3 succeeded where B2 failed on \BThreeOnlyBTwo{} tasks; B2 succeeded where B3 failed on \BTwoOnlyBThree{}; and \BThreeBTwoTies{} pairs tied.  The paired risk difference was \BThreeBTwoDifference.  The paired-bootstrap 95\% interval was $[\BThreeBTwoCILow,\BThreeBTwoCIHigh]$, and the exact two-sided McNemar value was $p=\BThreeBTwoP$.

The numerical interval excludes zero while the exact discordant-pair test does not reach a conventional 0.05 threshold.  That difference is unsurprising with only four discordant pairs and two distinct descriptive procedures.  More importantly, neither quantity has confirmatory interpretation after post-interim truncation.  The correct statement is that the reduced sample showed no observed B3-only wins and four B2-only wins, with a negative descriptive point estimate.

Worst/best assignment of the two null arms bounded the all-20-task B3--B2 difference from \BThreeBTwoNullLow{} to \BThreeBTwoNullHigh.  The estimate was \BThreeAfterBTwo{} when B3 followed B2 and \BThreeBeforeBTwo{} when B3 preceded B2.  The early and late launch-half estimates were \BThreeBTwoEarly{} and \BThreeBTwoLate, respectively.  The leave-one-task-out range was $[\BThreeBTwoLotoLow,\BThreeBTwoLotoHigh]$.  These diagnostics do not repair carryover or stopping bias, but they show how the descriptive estimate changes under the locked summaries.

\subsection{Secondary B3--B0 contrast}

Across \BThreeBZeroPairs{} valid pairs, B3-only successes numbered \BThreeOnlyBZero, B0-only successes \BZeroOnlyBThree, and ties \BThreeBZeroTies.  The paired difference was \BThreeBZeroDifference{} with interval $[\BThreeBZeroCILow,\BThreeBZeroCIHigh]$ and exact McNemar $p=\BThreeBZeroP$.  Null bounds were $[\BThreeBZeroNullLow,\BThreeBZeroNullHigh]$.

This comparison is secondary even before considering the truncation.  B0 was originally included only for the preselected 20-task subset, whereas B2 and B3 defined the planned 60-task principal contrast.  The observed negative B3--B0 difference is useful descriptive context, not a license to claim inferiority or to redesign the primary estimand after seeing data.

\begin{table*}[t]
\centering
\caption{Paired exploratory contrasts.  Intervals are paired-bootstrap percentile intervals; $p$ values are exact two-sided McNemar values.}
\label{tab:paired-results}
\small
\begin{tabular}{lrrrrrr}
\toprule
Contrast & Pairs & B3-only & Other-only & Ties & Difference & 95\% interval / $p$ \\
\midrule
B3--B2 & \BThreeBTwoPairs & \BThreeOnlyBTwo & \BTwoOnlyBThree & \BThreeBTwoTies & \BThreeBTwoDifference & $[\BThreeBTwoCILow,\BThreeBTwoCIHigh]$ / \BThreeBTwoP \\
B3--B0 & \BThreeBZeroPairs & \BThreeOnlyBZero & \BZeroOnlyBThree & \BThreeBZeroTies & \BThreeBZeroDifference & $[\BThreeBZeroCILow,\BThreeBZeroCIHigh]$ / \BThreeBZeroP \\
\bottomrule
\end{tabular}
\end{table*}

\subsection{Workflow closure}

None of the \BThreeClosureDenominator{} scored B3 cells reached \texttt{FINALIZATION\_ACKED}: \BThreeFinalized/\BThreeClosureDenominator.  The B3 apparatus-null cell has null closure.  Seven B3 artifacts nevertheless passed the blind Oracle.  Thus, workflow liveness and artifact efficacy empirically diverged.

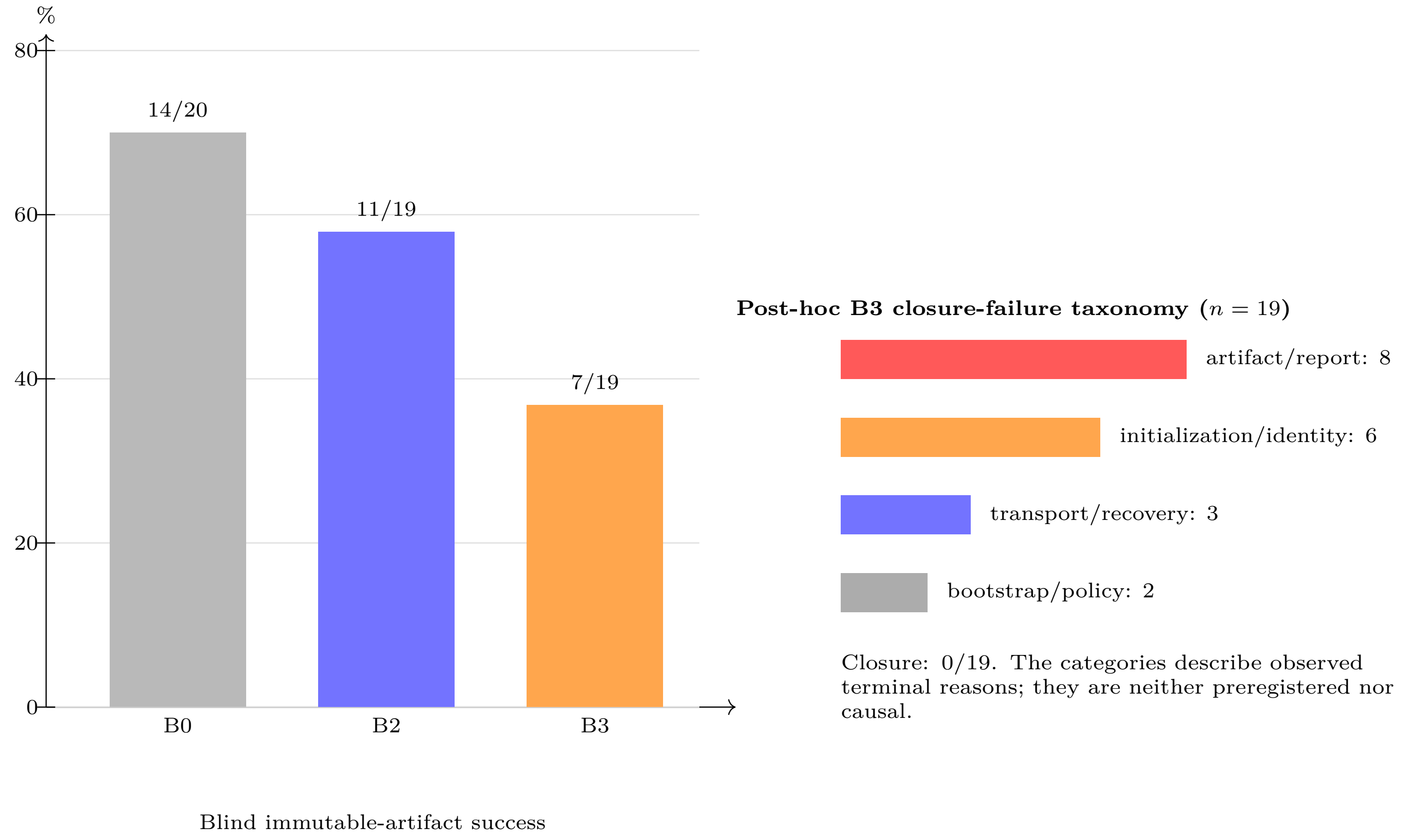
\begin{figure*}[t]
  \centering
  \resizebox{\linewidth}{!}{\begin{tikzpicture}[font=\scriptsize]
  \begin{scope}[x=0.105cm,y=0.095cm]
    \draw[->] (0,0) -- (0,82) node[above]{\%};
    \draw[->] (0,0) -- (76,0);
    \foreach \y in {0,20,40,60,80} {
      \draw (-1,\y) -- (1,\y) node[left=2pt]{\y};
      \draw[gray!25] (1,\y) -- (72,\y);
    }
    \fill[gray!55] (7,0) rectangle (22,\BZeroRate);
    \fill[blue!55] (30,0) rectangle (45,\BTwoRate);
    \fill[orange!70] (53,0) rectangle (68,\BThreeRate);
    \node[above] at (14.5,\BZeroRate) {\BZeroSuccesses/\BZeroScored};
    \node[above] at (37.5,\BTwoRate) {\BTwoSuccesses/\BTwoScored};
    \node[above] at (60.5,\BThreeRate) {\BThreeSuccesses/\BThreeScored};
    \node[below] at (14.5,0) {B0};
    \node[below] at (37.5,0) {B2};
    \node[below] at (60.5,0) {B3};
    \node[align=center] at (36,-14) {Blind immutable-artifact success};
  \end{scope}

  \begin{scope}[shift={(9.2cm,0)}]
    \node[font=\scriptsize\bfseries] at (2.0,4.6) {Post-hoc B3 closure-failure taxonomy ($n=19$)};
    \fill[red!65] (0,3.8) rectangle (4.0,4.25);
    \fill[orange!70] (0,2.9) rectangle (3.0,3.35);
    \fill[blue!55] (0,2.0) rectangle (1.5,2.45);
    \fill[gray!65] (0,1.1) rectangle (1.0,1.55);
    \node[right] at (4.1,4.03) {artifact/report: \FailureArtifactReport};
    \node[right] at (3.1,3.13) {initialization/identity: \FailureInitialization};
    \node[right] at (1.6,2.23) {transport/recovery: \FailureTransport};
    \node[right] at (1.1,1.33) {bootstrap/policy: \FailureBootstrap};
    \node[align=left, text width=64mm] at (3.2,.25) {Closure: \BThreeFinalized/\BThreeClosureDenominator. The categories describe observed terminal reasons; they are neither preregistered nor causal.};
  \end{scope}
\end{tikzpicture}}
  \caption{Left: blind artifact success by arm.  Right: post-hoc descriptive categories for the \BThreeClosureDenominator{} scored B3 closure failures.}
  \label{fig:results-failures}
\end{figure*}

Figure~\ref{fig:results-failures} places the two observations side by side.  B3 did produce some semantically correct artifacts, but the full lifecycle did not close in any scored B3 cell.  Counting those patches as workflow successes would weaken the state-machine definition after observing failure; counting them as Oracle failures would hide genuine task success.  We do neither.

\subsection{Outside-subset diagnostics}

Fourteen terminal cells completed before the administrative truncation but fell outside the selected population.  Six B2 cells included two Oracle successes, and eight B3 cells included three.  No outside-subset B0 cell was executed.  These cells remain visible as supplemental evidence, were not reclassified, and are not pooled into the reduced estimands.  Their existence is part of the stopping history, not a source of extra power for a post-hoc analysis.

\section{Operational Failure Analysis}

\subsection{Method and scope}

We analyze only the aggregate successor-010 reason inventory already present in the frozen public analysis.  We do not read, quote, or disclose raw task conversations, private test material, personal reports, or re-identifying artifacts.  The taxonomy was constructed after observing the terminal reasons.  It is descriptive, not preregistered; it assigns a mechanism-oriented label, not a definitive root cause.

The mapping is exhaustive for the \BThreeClosureDenominator{} scored B3 failures and excludes the one B3 apparatus-null cell.  Every non-null reason maps to exactly one of four categories.  The paper-local generator asserts the exact reason set and expected totals so that unmatched or duplicated reasons fail the build interface.

\subsection{Initialization and identity: \FailureInitialization{} failures}

This category includes Gateway initialization schema rejection, invalid initialization requests, invalid task receipt thread identity, and pre-dispatch thread-identity propagation timeout.  These failures occurred before a normal route/report lifecycle could close.  They illustrate why a Gateway cannot accept an unbound task merely because the App shows a plausible thread.  They also reveal that cross-layer identity propagation was too fragile for the fixed budget.

The architectural response is not to accept a guessed identifier.  It is to make task creation acknowledgement a machine operation with one durable request identity, exact-root bounded readback after ambiguous errors, and a runtime-generated registration request.  The model should decide which eligible Goal to route, not copy opaque App identifiers across schemas.

\subsection{Artifact and report contract: \FailureArtifactReport{} failures}

This largest category includes captured-diff report binding rejection, complete-diff reference algorithm mismatch, invalid runtime diff references, report-staging schema rejection, exhausted repair budgets, and an unresolved report-repair state.  In several cases the implementation artifact existed, but formal promotion failed because report identity did not match the runtime's canonical artifact representation.

The safety interpretation is bounded: the control plane refused to promote evidence that did not satisfy its identity contract.  The liveness interpretation is equally important: the official model-facing contract and runtime interface did not make the correct request reliably constructible.  Repeating stricter prose would retain the same failure mode.  Snapshot digests, canonical evidence paths, closed media types, and report-reference algorithms are deterministic values and should be derived by the runtime from handles rather than selected by the model.

\subsection{Transport and recovery: \FailureTransport{} failures}

Three failures remained in a sent-outbox recovery state until the repair budget expired.  The persistent outbox prevented blind replay, which is the intended safety property.  But the recovery path could not obtain a unique, admissible acknowledgement in time.  This category is the clearest expression of the safety--liveness tradeoff: refusing a duplicate protects identity while indefinite ambiguity prevents progress.

A better implementation would expose a single idempotent send primitive whose return value is durably linked to the prepared outbox.  Where the App cannot supply such an atomic primitive, the runtime should own bounded exact-target reconciliation and return a closed result to the model: adopted, absent and safe to send, duplicate and blocked, or unknown and paused.

\subsection{Bootstrap and policy: \FailureBootstrap{} failures}

One failure followed a Worker bootstrap protocol violation and one followed a forbidden private-path access before initialization.  The latter correctly preserved the experiment's privacy and Oracle boundary.  Neither observation implies that all policy violations are detected.  Together they show why the treatment cannot be permitted to broaden scope in order to rescue a cell.

\subsection{Cross-category interpretation}

The taxonomy does not show that the abstract state-machine idea is invalid.  It shows that the evaluated integration placed too much protocol adaptation on language-model roles.  The failures clustered at boundaries where opaque identifiers, digests, schema fields, or external task state had to be translated into exact requests.  That is a software architecture problem distinct from the model's ability to write a patch.

At the same time, these are not ignorable ``apparatus details'' inside B3.  The control plane is part of the evaluated end-to-end treatment, and all three arms shared the same 1,800-second outer budget.  A systems paper must report that the additional guarantees were accompanied by severe loss of normal completion capability.

\section{Discussion}

\subsection{What the safety evidence supports}

The aggregate execution record contains repeated terminal reasons labelled as stale, malformed, mismatched, duplicate, or unauthorized evidence.  Independently, source inspection shows explicit expected-version, artifact/report-binding, role-dispatch, outbox, and heartbeat precondition checks.  Together these sources establish inspectable enforcement points and observed rejection-labelled failures; they do not establish from the aggregate data that every rejection preserved canonical state or that no unsupported promotion was silently accepted.

This is a code-level and descriptive safety account, not a formal proof or an empirical false-accept rate.  The aggregate analysis contains neither request-level before/after canonical states nor a denominator of attempted invalid promotions.  A rejected request may also reflect a contract usability defect rather than a useful security boundary.  Future safety evaluation should prospectively enumerate attack and corruption classes, bind every attempt to privacy-safe before/after state receipts, and measure both false accepts and false rejects.

\subsection{Why liveness failed}

Liveness depends on more than the existence of a happy-path state diagram.  The model had to translate natural-language instructions into exact schema keys, choose among multiple digest identities, register App-created task identifiers, manage one outbox through ambiguous transport, and stage reports with canonical paths and media types.  Each boundary was individually strict, but the end-to-end probability of traversing all boundaries was low.  Zero finalizations are therefore not a small secondary blemish; they are evidence that the evaluated protocol was not operationally usable in this setting.

This result also explains why adding a Reviewer or a heartbeat is not free.  A role graph can make authority explicit and localize failure, but every edge adds a precondition and a possible stale identity.  LoopSkill's graph improved observability of where the workflow stopped.  The experiment does not show that the graph improved patch success.

\subsection{Efficacy under a fixed budget}

The common 1,800-second limit makes the comparison fair in one narrow sense: each end-to-end system receives the same wall-clock opportunity.  B3's lower artifact success cannot be dismissed by subtracting its protocol time after the fact.  The control plane is the treatment, so its overhead belongs in the measured effect.

The same limit narrows the research claim.  A 30-minute SWE-bench repair does not directly measure multi-hour autonomy, interruption recovery across days, evolving subgoals, or accumulated evidence over a long project.  Long-task evaluation work explicitly relates capability to task duration and reliability~\cite{kwa2025longtasks,starace2024rebench}; PaperBench decomposes much larger research-replication projects into many graded subtasks~\cite{starace2025paperbench}.  LoopSkill needs a separate prospective long-horizon study with tasks that actually require persistence, recovery, and multi-stage authority.  The present fixed-budget experiment cannot be retrospectively relabeled as that study.

\subsection{What should move from the model into runtime}

The model should retain semantic choices: interpret the issue, inspect code, propose an implementation, respond to substantive review, and decide among meaningful repair strategies.  The runtime should own deterministic protocol mechanics:

\begin{itemize}
  \item derive snapshot and report digests from opaque handles;
  \item generate canonical evidence paths and media types;
  \item fill state versions, lease, outbox, dispatch, and role identifiers;
  \item atomically acknowledge task creation or perform bounded exact-target reconciliation;
  \item expose one typed transition client generated from the same schema as the Gateway;
  \item pause and read back the registered heartbeat through a closed operation; and
  \item return structured recovery outcomes rather than asking the model to infer them from prose and logs.
\end{itemize}

This change preserves fail-closed verification while shrinking the surface on which a language model acts as a distributed transaction adapter.  The principle resembles secure supply-chain tooling: developers decide what to build, while tooling computes and verifies provenance metadata~\cite{torresarias2019intoto,slsa2026spec}.

\subsection{Implications for agent evaluation}

Agent evaluations often collapse progress into a final task score.  AgentBoard argues for fine-grained process measures~\cite{ma2024agentboard}, while LLM-as-a-judge work documents limitations of automated evaluation~\cite{zheng2023llmjudge}.  Our results show a complementary need: task effect, control-plane liveness, and evidence safety should be distinct reported dimensions.

For a governed agent, a useful scorecard should include at least: immutable-artifact correctness; percentage of eligible workflows reaching each major state edge; false acceptance and rejection under injected identity faults; recovery without duplicate side effects; time spent in semantic work versus protocol; and human interventions.  A single ``completed'' flag cannot distinguish a correct patch with failed closure from an incorrect patch with clean bookkeeping.

\section{Related Work}

\paragraph{Coding agents and software-engineering benchmarks.}
SWE-bench established executable evaluation on real GitHub issues~\cite{jimenez2024swebench}.  SWE-agent showed that agent-computer interface design changes repository-navigation and repair performance~\cite{yang2024sweagent}.  OpenHands provides a general platform for software-development agents~\cite{wang2025openhands}, while AutoCodeRover uses program structure and targeted search~\cite{zhang2024autocoderover}.  Agentless demonstrates that a simpler localization--repair--validation pipeline can be a strong and inexpensive baseline~\cite{xia2025agentless}.  LoopSkill is not another patch-generation method.  It is a lifecycle control layer whose efficacy cost must be evaluated alongside those simpler systems.

\paragraph{Interactive and long-horizon agent evaluation.}
AgentBench, GAIA, OSWorld, AgentBoard, and MLAgentBench evaluate multi-step agents across text, tools, operating systems, and experimentation~\cite{liu2024agentbench,mialon2024gaia,xie2024osworld,ma2024agentboard,huang2024mlagentbench}.  RE-Bench and task-completion time-horizon work emphasize difficult research-engineering tasks and reliability over increasing human task duration~\cite{starace2024rebench,kwa2025longtasks}.  PaperBench evaluates full research replication with hierarchical rubrics~\cite{starace2025paperbench}.  These works motivate richer horizons and process measurement.  Our current experiment remains a fixed 30-minute coding study and therefore cannot claim the long-horizon coverage of those benchmarks.

\paragraph{Agent loops, self-correction, and review.}
ReAct interleaves reasoning and action~\cite{yao2023react}; Toolformer studies learned API use~\cite{schick2023toolformer}; Reflexion and Self-Refine use feedback to improve later attempts~\cite{shinn2023reflexion,madaan2023selfrefine}.  Loop engineering describes reusable trigger--goal--verification--stopping structures~\cite{macedo2026loops}.  Proof-or-Stop treats lifecycle advancement as evidence-gated~\cite{huang2026prooforstop}, and commit-time authorization focuses on freshness and binding at the durability boundary~\cite{santosgrueiro2026authority}.  LoopSkill combines those concerns with explicit canonical state and original-outbox recovery.  Its independent Reviewer is a maker--checker boundary, not a guarantee that model judgment is unbiased; LLM-judge studies caution against that stronger inference~\cite{zheng2023llmjudge}.

\paragraph{Durable workflows and state machines.}
State-machine replication establishes ordered, deterministic service transitions~\cite{schneider1990statemachine,lamport1998parliament}.  Sagas decompose long transactions and make partial failure explicit~\cite{garcia1987sagas}; workflow patterns catalogue reusable control-flow structures~\cite{vanderaalst2003workflow}.  Reliable State Machines, Netherite, and Temporal address durable execution across process failures~\cite{burckhardt2019rsm,burckhardt2021netherite,temporal2026docs}.  The transactional outbox links durable intent to an external message~\cite{aws2026outbox}.  LoopSkill adapts these ideas to model-driven coding work, where a role can generate plausible but structurally wrong protocol data and where external artifact correctness remains separate from workflow state.

\paragraph{Provenance and attestation.}
W3C PROV models entities, activities, and agents that produce data~\cite{w3c2013prov}.  in-toto binds supply-chain steps and functionaries to artifacts~\cite{torresarias2019intoto}; SLSA standardizes build provenance levels and predicates~\cite{slsa2026spec}; software-update security work shows why freshness, role separation, and complete repository views matter~\cite{cappos2010tuf}.  LoopSkill borrows the insistence on identity-bound evidence but does not claim cryptographic supply-chain security.  Its host observations and local digests are operational evidence under a stated trust model.

\section{Threats to Validity and Evidence Ethics}

\subsection{Internal validity}

The dominant threat is post-interim administrative truncation.  Although the reduced tasks came from a previously selected outcome-blind subset, the choice to stop at that subset was made after outcomes were visible.  Confirmatory interpretation is invalid.  Condition order was deterministic rather than randomized at execution time, and platform memory remained enabled.  Prior same-task exposure, launch order, and memory can create carryover.  Order strata and leave-one-out summaries describe but do not remove this confounding.

The host exposed no controllable model random seed.  Pairing reduces task heterogeneity but does not eliminate stochastic execution variance.  One null occurred in B2 and one in B3, changing complete-pair denominators.  Worst/best bounds show the limited range of possible all-task differences but do not recover the missing outcomes.  The two nulls were not rerun, which protects against selective rescue at the cost of precision.

\subsection{Construct validity}

Blind test success measures whether the frozen patch passes prespecified tests; it is not a complete measure of maintainability, security, or developer value.  \texttt{FINALIZATION\_ACKED} measures conformity to the evaluated lifecycle protocol, not semantic correctness.  The post-hoc reason taxonomy reflects labels emitted by that protocol and may merge distinct root causes.  The cost construct is incomplete because the locked public analysis does not include a complete token or monetary ledger.

The 1,800-second budget measures fixed-budget end-to-end efficacy.  It may overemphasize B3's startup cost relative to a multi-hour task, while longer tasks may introduce additional failure opportunities.  No direction can be assumed without a new experiment.

\subsection{External validity}

The study covers one selected SWE-bench Lite subset, one model/host setup, enabled platform memory, one local LoopSkill version, and one Desktop integration.  Qualification narrowed the population to locally supported images.  Results do not establish performance on other languages, repositories, providers, organizations, or truly long-horizon projects.  The title's lifecycle framing describes the system's intended problem domain, not a demonstrated long-horizon advantage.

\subsection{Self-application and observer independence}

LoopSkill was used only inside B3.  Treatment-external code selected and scheduled cells, froze artifacts, ran the Oracle, classified results, locked analysis, and generated manuscript numbers.  This limits direct self-governance but cannot remove all shared-platform effects.  The independent manuscript review checks evidence boundaries and implementation descriptions; it is not an independent replication of the experiment.

\subsection{Privacy and public evidence}

The paper uses aggregate study results, public task metadata, and implementation descriptions.  Raw personal reports, private conversations, hidden Oracle content, and re-identifying artifacts are excluded.  The public bundle is allowlist-only and contains no raw task transcripts, hidden tests, or private Oracle material.

\paragraph{Artifact availability.} \PublicationAvailabilityStatement

\paragraph{AI assistance disclosure.} OpenAI Codex assisted with drafting, implementation support, and local validation.  The named human authors reviewed the manuscript and retain responsibility for its claims, evidence boundary, licensing, and publication decisions; the system is not listed as an author.

\balance
\section{Conclusion}

LoopSkill treats coding-agent lifecycle advancement as a transaction over current identities and evidence rather than a conversational claim.  Its State Gateway, durable outbox, runtime-mediated artifact/report staging, independent review route, original-outbox recovery, and heartbeat-gated finalization make authority and failure boundaries explicit.

The exploratory evaluation shows the cost of the evaluated integration as clearly as its controls.  In the administratively truncated 20-task population, B3 produced \BThreeSuccesses{} correct artifacts among \BThreeScored{} scored cells, compared with \BTwoSuccesses/\BTwoScored{} for B2 and \BZeroSuccesses/\BZeroScored{} for B0.  The principal paired estimate was negative, and no scored B3 workflow reached finalization.  Aggregate terminal reasons localized observed failures, while source inspection identified the checks intended to reject unsupported promotion; the available aggregate evidence does not establish a false-accept rate or request-level no-advance result.

The responsible conclusion is a safety--liveness tradeoff, not a superiority claim and not a blanket claim of ineffectiveness.  The evaluated control plane enforced strong boundaries but required the language model to perform too much deterministic protocol adaptation, producing severe liveness loss within a 1,800-second budget.  The next design should retain identity checks while moving schema construction, digest derivation, task acknowledgement, and recovery reconciliation into typed runtime operations.  A new prospectively frozen long-horizon study is required to test whether that architecture improves reliable multi-stage work without repeating the present fixed-budget cost.
\label{body-end}

\clearpage
\bibliographystyle{abbrvnat}
\bibliography{full-manuscript-references}

\clearpage
\appendix

\section{Canonical Transition Summary}

Table~\ref{tab:transition-summary} summarizes the main protocol edges.  It is explanatory rather than a substitute for the versioned machine schema.

\begin{table*}[t]
\centering
\small
\caption{Principal lifecycle operations and required evidence.}
\label{tab:transition-summary}
\begin{tabularx}{\textwidth}{@{}P{35mm}P{32mm}YY@{}}
\toprule
Operation & Authorized caller & Required current binding & Advancement / rejection \\
\midrule
\texttt{PREPARE\_ROUTE} & Controller via Gateway & State version, Goal, lease, role, target, payload & Creates one prepared outbox; mismatch leaves state unchanged \\
\texttt{RECORD\_ROUTE\_SENT} & Controller via Gateway & Prepared outbox and exact App target observation & Marks the same route sent; ambiguous target cannot be guessed \\
\texttt{STAGE\_REPORT} & Dispatched Worker or Reviewer via codec & Outbox, role dispatch, registered worktree, artifact/report/evidence bytes & Creates immutable staged handle; invalid schema or digest produces no ACK \\
\texttt{ACK\_ROUTE\_RESULT} & Controller via Gateway & Sent outbox and matching staged report & Accepts the result predecessor; foreign report is rejected \\
\texttt{REPORT\_RECOVERY} & Controller via Gateway & Original outbox and persistent staged identity & Returns original handle; never reruns product work \\
\texttt{ACK\_ROUTE\_RESULT} (review route) & Controller via Gateway & Sent review outbox, Worker predecessor, staged Reviewer report & Accepts the review-route result; runtime internally projects the review decision \\
\texttt{PREPARE\_FINALIZATION} & Controller via Gateway & Complete required routes and reviews & Creates nonterminal finalization outbox \\
\texttt{ACK\_FINALIZATION} & Controller via Gateway & Finalization outbox, terminal predecessor, exact paused heartbeat readback & Produces internal \texttt{FINALIZATION\_ACKED}; never public authority \\
\bottomrule
\end{tabularx}
\end{table*}

\section{Identity-Binding Invariants}

The following checklist defines the manuscript-level model of an admissible Worker result:

\begin{enumerate}
  \item The request's expected version equals the current canonical version.
  \item The caller role is registered for the operation and target.
  \item Goal, milestone, lease, and roadmap identities equal those in canonical state.
  \item The outbox is the unique active route for that role and Goal.
  \item The target task and observed workspace root match the registered route.
  \item The received dispatch payload digest matches the prepared payload.
  \item Artifact identity is derived from the complete authorized post-work snapshot.
  \item Evidence-file digests bind exact source bytes and canonical staged paths.
  \item Formal report bytes bind the same outbox, artifact, result, and validation claims.
  \item A Reviewer uses a distinct review dispatch and does not mutate the Worker artifact.
  \item Finalization binds the accepted result/review predecessor and the exact registered heartbeat's paused readback.
  \item Internal terminal state confers no authority for commit, push, release, publication, or submission.
\end{enumerate}

The implementation may enforce additional version-specific constraints.  If prose and machine schema disagree, the evaluated schema and runtime at the frozen commit are authoritative for implementation claims.

\section{Complete Sensitivity and Diagnostic Results}

\begin{table*}[t]
\centering
\caption{Locked descriptive sensitivity summaries.}
\label{tab:sensitivity}
\small
\begin{tabularx}{\textwidth}{@{}P{34mm}P{28mm}YY@{}}
\toprule
Diagnostic & B3--B2 & B3--B0 & Interpretation boundary \\
\midrule
Valid complete pairs & \BThreeBTwoPairs & \BThreeBZeroPairs & Null arms are excluded only from complete-pair estimates \\
Worst/best null bounds & $[\BThreeBTwoNullLow,\BThreeBTwoNullHigh]$ & $[\BThreeBZeroNullLow,\BThreeBZeroNullHigh]$ & Assigns missing outcomes in adverse/favorable directions over all 20 tasks \\
B3 after comparison & \BThreeAfterBTwo{} ($n=9$) & \BThreeAfterBZero{} ($n=9$) & Descriptive condition-order stratum \\
B3 before comparison & \BThreeBeforeBTwo{} ($n=9$) & \BThreeBeforeBZero{} ($n=10$) & Does not remove memory carryover \\
Early launch half & \BThreeBTwoEarly & \BThreeBZeroEarly & Diagnostic only \\
Late launch half & \BThreeBTwoLate & \BThreeBZeroLate & Diagnostic only \\
Leave-one-task-out range & $[\BThreeBTwoLotoLow,\BThreeBTwoLotoHigh]$ & $[\BThreeBZeroLotoLow,\BThreeBZeroLotoHigh]$ & Descriptive influence range \\
\bottomrule
\end{tabularx}
\end{table*}

The bootstrap used 100,000 paired replicates.  Locked seeds were 3,312,011 for B3--B2 and 3,312,012 for B3--B0.  Bootstrap intervals and exact McNemar values answer different descriptive questions and neither restores a confirmatory design after the sample was administratively truncated.

\section{Post-hoc Failure Mapping}

Table~\ref{tab:failure-map} lists the exact aggregate reason-to-category mapping.  Counts are read from the frozen public analysis.  Long reason strings are protocol labels, not quotations from private task content.

\begin{table*}[t]
\centering
\scriptsize
\caption{Exhaustive mapping for the 19 scored B3 closure failures.}
\label{tab:failure-map}
\begin{tabularx}{\textwidth}{@{}P{39mm}YP{9mm}@{}}
\toprule
Category & Aggregate terminal reason & Count \\
\midrule
Initialization/identity & \texttt{BLOCKED\_GATEWAY\_INITIALIZE\_SCHEMA\_REJECTED} & 1 \\
& \texttt{BLOCKED\_GATEWAY\_INITIALIZE\_SCHEMA\_REJECTED\_TWICE} & 1 \\
& \texttt{BLOCKED\_STATE\_GATEWAY\_INITIALIZE\_REQUEST\_INVALID} & 1 \\
& \texttt{BLOCKED\_STATE\_GATEWAY\_TASK\_RECEIPT\_INVALID\_THREAD\_ID} & 1 \\
& \texttt{THREAD\_IDENTITY\_PROPAGATION\_TIMEOUT\_PRE\_DISPATCH} & 2 \\
\midrule
Artifact/report contract & \texttt{BLOCKED\_CAPTURED\_GIT\_DIFF\_REPORT\_BINDING\_REJECTED\_REPAIR\_EXHAUSTED} & 1 \\
& \texttt{BLOCKED\_COMPLETE\_DIFF\_REFERENCE\_ALGORITHM\_MISMATCH} & 2 \\
& \texttt{BLOCKED\_RUNTIME\_CODEC\_COMPLETE\_DIFF\_REFERENCE\_INVALID} & 1 \\
& \texttt{BLOCKED\_STAGE\_REPORT\_COMPLETE\_DIFF\_REFERENCE\_SCHEMA\_REJECTED} & 1 \\
& \texttt{REPAIR\_BUDGET\_EXHAUSTED\_RUNTIME\_DIFF\_REFERENCE\_NOT\_CLOSED} & 1 \\
& \texttt{REPAIR\_REPORT\_STAGING\_ALGORITHM\_MISMATCH\_BUDGET\_EXPIRED\_UNRESOLVED\_SENT\_OUTBOX} & 1 \\
& \texttt{REPAIR\_REQUIRED} & 1 \\
\midrule
Transport/recovery & \texttt{WAITING\_TRANSPORT\_RECOVERY\_BUDGET\_EXPIRED\_UNRESOLVED\_SENT\_OUTBOX} & 3 \\
\midrule
Bootstrap/policy & \texttt{E2E\_PROTOCOL\_VIOLATION\_WORKER\_BOOTSTRAP\_SYSTEM\_ERROR} & 1 \\
& \texttt{FORBIDDEN\_PRIVATE\_PATH\_ACCESS\_PREINITIALIZATION\_POLICY\_STOP} & 1 \\
\bottomrule
\end{tabularx}
\end{table*}

Category totals are initialization/identity \FailureInitialization, artifact/report contract \FailureArtifactReport, transport/recovery \FailureTransport, and bootstrap/policy \FailureBootstrap.  The mapping is a post-hoc descriptive lens.  It must not be used as a preregistered causal decomposition or to selectively remove B3 failures from the efficacy result.

\section{Reproduction and Public-Evidence Boundary}

The manuscript's numerical interface can be reproduced without accessing hidden tests or raw cell content:

\begin{enumerate}
  \item Verify the frozen public analysis JSON against its expected SHA-256.
  \item Run the paper-local result-asset generator.  It asserts cell accounting, arm summaries, pair counts, intervals, null bounds, closure counts, and failure categories.
  \item Run the generator in check mode to confirm that the committed LaTeX macro file is byte-identical to a fresh rendering.
  \item Build the manuscript with Tectonic from the paper directory.
  \item Run the publication-bundle validator, which verifies the aggregate-result digest, generated macros, manuscript source, release assets, public allowlist, licenses, and absence of prohibited local/private material without executing a hidden Oracle.
\end{enumerate}

The public-evidence boundary permits the aggregate result JSON, result-interface validator, claim--evidence table, implementation description, and privacy-screened manuscript assets.  It excludes hidden test content, private Oracle inputs, raw task transcripts, personal reports, secrets, local absolute paths, and evidence not explicitly allowlisted.  The public aggregate preserves the locked descriptive results but is not an independent replication of the hidden Oracle.

\end{document}